\documentclass[%
 superscriptaddress,reprint,
 amsmath,amssymb,
 aps,
]{revtex4-2}

\usepackage{graphicx}
\usepackage{dcolumn}
\usepackage{bm}

\newcommand{\appropto}{\mathrel{\vcenter{
  \offinterlineskip\halign{\hfil$##$\cr
    \propto\cr\noalign{\kern2pt}\sim\cr\noalign{\kern-2pt}}}}}

\newcommand{\bra}[1]{\langle #1 |}
\newcommand{\ket}[1]{| #1 \rangle}

\begin{document}

\preprint{APS/123-QED}

\title{Unambiguous discrimination of high harmonic generation mechanisms in solids}
\author{Graham G. Brown}
\email{brown@mbi-berlin.de}
\affiliation{Max Born Institute, Max-Born-Stra\ss e 2A, 12489, Berlin, Germany}
\author{\'{A}lvaro Jim\'{e}nez-Gal\'{a}n}
\affiliation{Max Born Institute, Max-Born-Stra\ss e 2A, 12489, Berlin, Germany}
\affiliation{Joint Attosecond Science Laboratory, National Research Council of Canada and University of Ottawa, Ottawa, Canada}
\author{Rui E. F. Silva}
\affiliation{Max Born Institute, Max-Born-Stra\ss e 2A, 12489, Berlin, Germany}
\affiliation{ICMM, Centro Superior de Investigaciones Científicas, Madrid, Spain}
\author{Eleftherios Goulielmakis}
\affiliation{Institute for Physics, University of Rostock  
Albert-Einstein-Str. 23
18059 Rostock}

\author{Misha Ivanov}%
\affiliation{Max Born Institute, Max-Born-Stra\ss e 2A, 12489, Berlin, Germany}
\affiliation{Department of Physics, Humboldt University, Newtonstra\ss e 15, 12489 Berlin, Germany}
\affiliation{Blackett Laboratory, Imperial College London, London SW7 2AZ, United Kingdom}

\date{\today}

\begin{abstract}
Using real-space view of high harmonic generation (HHG) in solids, we develop a physically transparent and gauge-invariant approach for distinguishing intraband and interband HHG mechanisms. Our approach relies on resolving the harmonic emission according to the separation between Wannier states involved in radiative transitions. We show that the intra- and interband HHG emission exhibit striking qualitative differences in their dependence on this separation and can be clearly distinguished using the Wannier basis.
\end{abstract}

\maketitle


The extension of high harmonic generation (HHG) spectroscopy from the gas-phase to condensed matter has led to many important innovations and discoveries. These include novel spectroscopy of field-free \cite{allOpticalBandStructure} and laser-driven electronic band structure \cite{Uzan-Narovlansky2022}, topological systems \cite{HHGTPT,Baykusheva2021,PhysRevLett.120.177401}, strongly correlated systems \cite{hhgManyBody,PhysRevA.105.053118}, and Berry curvature \cite{HHGBC1}, to name but a few examples. From a technological perspective, HHG in solids offers novel routes to compact  solid-state sources of bright, coherent, ultrashort pulses in the deep and extreme UV regions \cite{Korobenko:21}. 

\begin{figure*}[t]
	\centering
	\includegraphics[width=\textwidth]{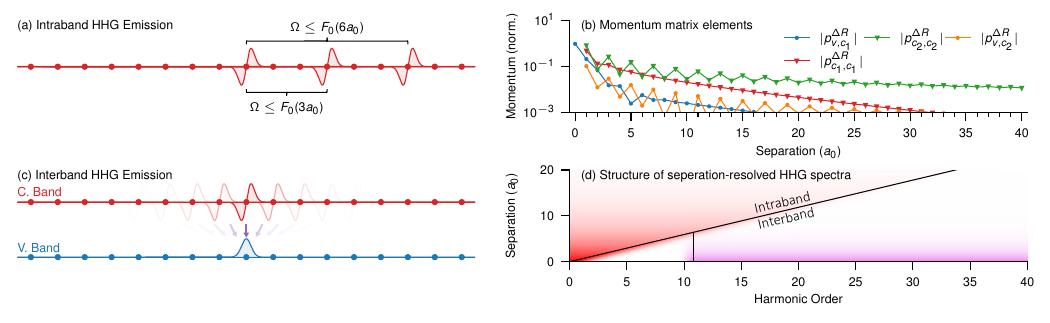}
	\caption{(a) In the Wannier basis, the photon energy $\Omega$ of intraband HHG emission is constrained by the product of the peak electric field amplitude $F_0$ and the separation $\Delta R$ between the Wannier states involved in the transition. The range of photon energies related to separations of three and six lattice sites are overlaid. (b) The transition cross-section for interband (circles) rapidly decays with electron-hole separation \cite{WQC} and is greatest for small separations, whereas the intraband transition cross-section (triangles) is much greater for large separations. (c) Thus, interband emission is dominated by small electron-hole separations, shown by fading opacity of conduction band Wannier functions. (d) These mechanisms result in a distinct structure when HHG spectra are resolved according to separation, wherein intraband (interband) emission occurs in the red (purple) regions. The oblique and vertical lines denote the intraband described by Eq. (\ref{eq:intrabandDR}) and the harmonic order of the minimum band gap.}	
	\label{fig:1}
\end{figure*}

Just like gas-phase HHG \cite{PhysRevLett.71.1994,PhysRevA.49.2117}, HHG in solids is often described as a combination of three steps \cite{vampa1}: (1) optical tunnelling promotes an electron to the conduction band and leaves a hole in the valence band, (2) the electron-hole pair is accelerated by the strong field, and (3) the electron-hole pair may recombine, emitting light. However, in contrast to the gas-phase, where high harmonic emission  occurs only during the final recombination step \cite{PhysRevA.49.2117}, in solids the high harmonic emission can occur during each of the three steps. In particular, it may arise from injection currents during tunnelling \cite{Jurgens2020}, the motion of the electron-hole pair during propagation (known as intraband harmonic emission) \cite{firstHHGSolid,Goulielmakis2022}, and during the final recombination step (known as interband emission) \cite{vampa1}. The latter two mechanisms tend to dominate experimental spectra for sufficiently high harmonic orders.

It is generally believed that intra- and interband HHG exhibit distinct spectra-temporal 
features \cite{vampa1}: intraband HHG emission occurs virtually simultaneously at all frequencies, while the interband emission carries characteristic chirp, which maps the electronic structure and sub-cycle dynamics of the electron-hole pair onto the spectral phase of interband HHG spectra. However, this analysis leads to clear conclusions only if one of the mechanisms dominates. Due to distinctly different spectro-temporal characteristics, the discrimination between intra- and interband HHG in either an experiment or numerical simulation is important for both qualitative and quantitative interpretation of experiments, especially when high harmonics are used as a spectroscopic tool to follow attosecond electron dynamics in solids.

One route towards discriminating between intra- and interband HHG emission numerically is to 
project the dynamics on the field-free Bloch basis at 
the instantaneous crystal momenta  $\mathbf{k}(t)=\mathbf{k}+\mathbf{A}(t)$, where $\mathbf{A}(t)$ is the field vector potential, separating coherences between the bands from the  intra-band current. Conceptually, the drawback of this approach is that the field-free states of the system are hardly relevant in the presence of a strong electric field which should significantly alter the band structure. Thus, what appears as inter-band coherences, may simply reflect the light-induced polarization of the bands. To address this problem, one can attempt to use the instantaneous eigenstates of the system \cite{Yue:22,PhysRevB.98.235202}, which include the electric field in the quasistatic approximation, treating time as a parameter. This, however, is only applicable for small driving field frequencies. Moreover, for more complex systems, significant numerical issues quickly arise, leading to numerical instabilities \cite{Yue:22}. Regardless of the approach, such formal discriminations between intra- and interband HHG emission introduces significant computational overhead and is neither unique nor gauge-invariant \cite{PhysRevA.91.013405}, making the relative role of the two key mechanisms a highly controversial issue \cite{PhysRevA.92.033845,PhysRevLett.116.016601,Garg2016,Luu2015}.

Here we demonstrate an approach which appears to resolve all the difficulties mentioned above, carries no additional computational cost, and is both gauge-invariant and unambiguous. We find that inter- and intraband HHG emission exhibit strikingly different properties when analyzed in real space. To this end, we describe solid-state HHG in real space using the Wannier basis \cite{wannierOG}, which describes periodic systems using a set of electronic orbitals localized around individual lattice sites, as opposed to the inherently delocalized Bloch basis. The real-space perspective afforded by the Wannier basis allows for the the resolution of radiative transitions according to the separation within a lattice between different Wannier states involved \cite{WQC,wannierHHGLewenstein,intrabandWannier}. 

In real-space, the dynamics leading to intra- and interband HHG exhibit distinct differences. The interband HHG is highly reminiscent of gas-phase HHG, wherein an electron-hole pair is accelerated in space and recombines when their separations are small. In contrast, we find that energy conservation requires that intraband HHG emission at high harmonic orders necessarily involves transitions between Wannier states with large spatial separations. These distinct dependences on the separation between Wannier states allows for unambiguous discrimination between the emission mechanisms.

Let us start with the semiclassical picture of solid-state HHG expressed in the Wannier basis \cite{intrabandWannier,WQC}. Applying the saddle-point approximation to solve for the intraband HHG spectrum yields the following constraint on the maximum intraband photon emission $\Omega$ with respect to the peak driving electric field amplitude $F_0$ and the separation between Wannier states $\Delta R$ involved in a radiative intraband transition \cite{intrabandWannier}:

\begin{equation}
	\Omega \le F_0 \Delta R .
	\label{eq:intrabandDR}
\end{equation}

\noindent 
This is qualitatively depicted in Fig. \ref{fig:1} (a), which shows three Wannier functions in the conduction band situated at different lattice sites (red dots). Intraband transitions between these states lead to photon emission at frequencies $\Omega$ which are constrained to be less than the product between their separation and the peak driving electric field amplitude, as indicated by the overlaid text. For typical driving fields, intraband emission at frequencies above the minimum band gap will arise only from transitions across large separations.

The energy of interband emission, on the other hand, is related to the coherent dynamics of the electron-hole wave packet and the cross-section for interband transitions which rapidly decays with the separation between two Wannier states \cite{PhysRevB.100.195201,RevModPhys.84.1419}. This is depicted in (b), where the interband (intraband) momentum matrix element magnitudes of the system used for our numerical simulations are shown by the circular (triangular) markers. Thus, interband HHG emission will be most efficient for transitions between Wannier states with small $| \Delta R |$, as depicted by the diminishing opacity of conduction-band Wannier functions in Fig. \ref{fig:1} (c). These distinct dependencies of the real-space generation mechanisms of intra- and interband HHG on the separation between Wannier states results in a clear structure when the emission is resolved according to $\Delta R$. This is qualitatively depicted in (d), where the red (purple) regions denote the separations where intraband (interband) HHG emission is expected to occur at various harmonic orders based on the parameters we use for our numerical simulations below. The vertical black line denotes the harmonic order of the minimum band gap and the oblique line depicts the intraband cutoff frequency described by Eq. (\ref{eq:intrabandDR}).

\begin{figure*}[ht]
	\includegraphics[width=\textwidth]{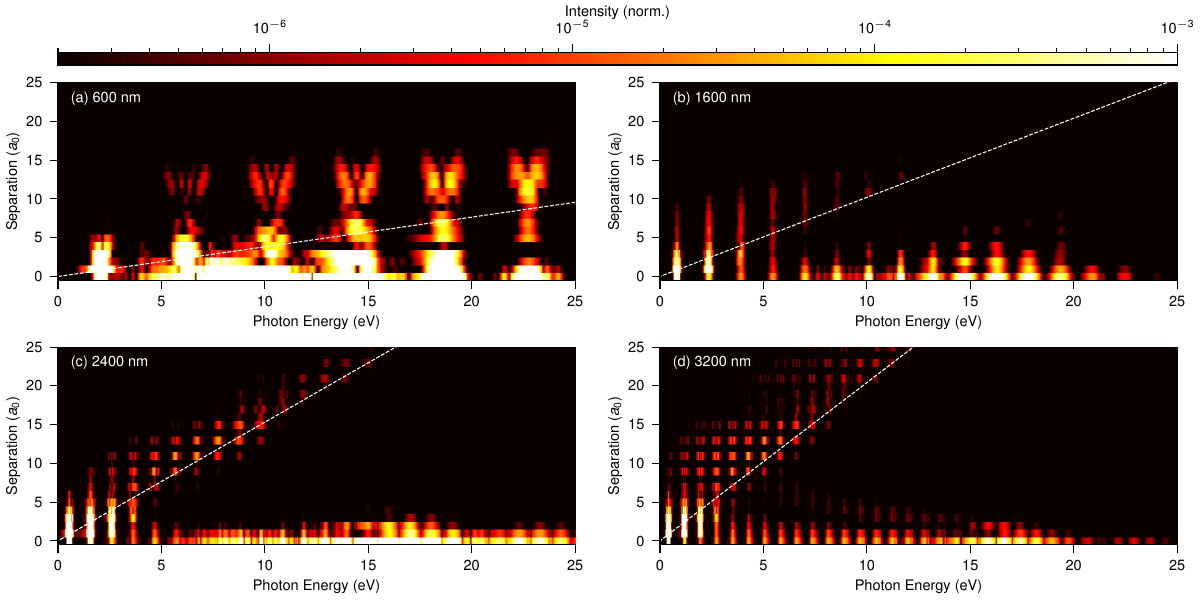}
	\caption{The lattice site separation-resolved total HHG spectra generated using driving lasers with wavelengths (a) 600 nm, 1600 nm, (c) 2400 nm, and (d) 3200 nm. All spectra are calculated with a dephasing time of 10 fs using the formalism presented in \cite{PhysRevB.97.144302}, one valence and seven conduction bands. The driving laser is a 5 cycle Gaussian pulse with an intensity set such that all simulations have an equal peak vector potential amplitude, such that the peak intensity for the 3200 nm case is $3.1 \times 10^{11}$ W/cm$^{2}$. In all figures, the vertical green line and oblique white line denote the harmonic order of the minimum band gap and the intraband cutoff for each separation, respectively.}
	\label{fig:2}
\end{figure*}
 
We now confirm these predictions through numerical simulations of solid-state HHG. We consider a one-dimensional periodic system with lattice constant $a_0 = 8$ a.u. and periodic Mathieu-type potential $v(x)$ with strength $0.37$ a.u. \cite{hhgFromBEiS} for simplicity and our model formulated in the Wannier basis is described in a prior work \cite{firstHHGSolid}. Our system is described through a set of Wannier states, $\ket{w_{m, R}}$, representing an electron orbital in band $m$ localized about the lattice site at position $R = n a_0$ (integer $n$). Generally, both the Hamiltonian and momentum operators in the Wannier basis couple Wannier states situated at different lattice sites $R$ and $R'$ with separation $\Delta R = R - R'$. All required operators can be expressed as a summation over contributions corresponding to the coupling between Wannier states for each possible separation $\Delta R$. The  component of the momentum operator $\hat{p}$, that most relevant for this work, coupling states separated by $\Delta R$ is

\begin{equation}
	\hat{p}_{\Delta R} = \sum_{m, m'} \sum_{R} \tilde{p}_{m, m'}^{\Delta R} \ket{\phi_{m, R}} \bra{\phi_{m', R - \Delta R}}, 
	\label{eq:wannierMomentumDR}
\end{equation}

\noindent 
where $\tilde{p}_{m, m'}^{\Delta R}$ is the Fourier transform of the respective Bloch basis momentum matrix elements \cite{firstHHGSolid} and the total momentum operator is obtained by summing Eq. (\ref{eq:wannierMomentumDR}) over all $\Delta R$.

We simulate the interaction of the system with an external field using the density matrix formalism \cite{firstHHGSolid} and describe the interaction with the electromagnetic field using the length and velocity gauges \cite{hhgFromBEiS}. After solving for the time-dependent density matrix, the total current associated with radiative transitions between states separated by a distance $\Delta R$, $j_{\Delta R}(t)$, is calculated as the trace of the product of $\hat{p}_{\Delta R}$ in Eq. (\ref{eq:wannierMomentumDR}) with the density matrix, from which we calculate the HHG emission arising from transitions separated by the absolute distance $| \Delta R |$ by summing the spectra obtained from $j_{\Delta R}(t)$ and $j_{-\Delta R}(t)$. The total HHG spectrum is then given by coherently summing the spectra calculated for all separations.

Fig. \ref{fig:2} shows the separation-resolved HHG spectra generated by a five-cycle Gaussian pulse with driving field wavelengths of (a) 600 nm, (b) 1600 nm, (c) 2400 nm, and (d) 3200 nm. The peak intensity for each simulation is chosen so that the peak vector potential is equal for all simulations, defined through the peak intensity of $3.1 \times 10^{11}$ W/cm$^{2}$ for the simulation using a 3200 nm wavelength. For all simulations, we include one valence and seven conduction bands. The dashed green and white lines in Fig. \ref{fig:2} denote the harmonic order of the minimum band gap and the cutoff for intraband HHG emission for each separation calculated using Eq. (\ref{eq:intrabandDR}) as depicted in Fig. \ref{fig:1} (d), respectively.


The structure of the spectra shown in Figs. \ref{fig:2} closely resemble the predictions of the semiclassical model in Fig. \ref{fig:1} (d) and this becomes increasingly apparent as the driving laser wavelength is increased. In particular, within the regions we have labelled as intraband in Fig. \ref{fig:1} (d), the HHG spectra in Fig. \ref{fig:2} exhibit a clear linear relationship between harmonic order and lattice site separation, whereas the spectra exhibit a prominent plateau structure until the HHG cutoff near harmonic order 35 in the interband region. The staggered structure of the intraband emission with separation is due to the magnitude of the intraband momentum matrix elements, as depicted in Fig. \ref{fig:1} (c).

To confirm the relationship between the HHG mechanism and separation between Wannier states, we depict the separation-resolved inter- and intraband HHG spectra obtained directly from the simulation performed using the length gauge, for which one can use discrimination according to band index \cite{PhysRevB.100.195201,RevModPhys.84.1419}. The separation-resolved inter- and intraband HHG spectra are shown in Figs. \ref{fig:2} (c) and (d), respectively. The relationship between inter- and intraband HHG emission distinguished using the band index and the distance between the Wannier states is unambiguous and clearly reflects the structure predicted by our semiclassical analysis. 

\begin{figure}[t]
	\centering
	\includegraphics[width=\columnwidth]{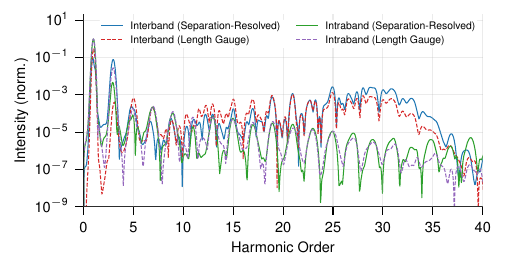}
	\caption{The (dashed red) interband and (dashed purple) intraband HHG spectra calculated from the length gauge simulation presented in Fig. \ref{fig:2} (a) and the (solid blue) interband and (solid green) intraband HHG spectra obtained from the discrimination according to the separation between Wannier states obtained by coherently summing the spectra in regions ($i$) and ($ii$), respectively.}	
	\label{fig:3}
\end{figure}

Finally, we present the principle finding of this work and show how the inter- and intraband HHG spectra can be obtained solely from the total separation-resolved HHG spectrum. We do this for the spectrum calculated using the length gauge in Fig. \ref{fig:2} (a) by coherently summing the spectra within the regions labelled as $(i)$ and $(ii)$, corresponding to the regions labelled as intra- and interband in Fig. \ref{fig:1} (a). This is depicted in Fig. \ref{fig:3}, where the inter- and intraband spectra obtained by our separation-resolved discrimination of the length gauge spectra from Fig. \ref{fig:2} (a) are depicted by the solid blue and green lines, respectively. For comparison, we also show the inter- and intraband HHG spectra obtained from the band-resolved discrimination of the spectra in Fig. \ref{fig:2} (a) by the dashed red and purple lines, respectively. While the two methods generally agree, there are also differences, visible especially beyond harmonic 27. The disagreement reflects the ambiguity in discriminating the inter- and intra-band mechanisms using the band index, especially as the number of involved conduction bands increases \cite{PhysRevA.91.013405}, the latter becoming more significant with the growing hamonic order. Our approach distinguishes between two distinct HHG generation mechanisms whose observable spectral characteristics are innately related to and distinguished by their underlying real-space dynamics and depends only on the total current, a gauge-invariant quantity. 

Instead of depending on the choice of electromagnetic gauge, our approach depends on the choice of gauge to describe the phase of the Bloch states, which are used to calculate the Wannier basis \cite{wannierOG,kohnWannier,RevModPhys.84.1419}. In this sense, we have traded one gauge-dependence for another. This gauge-dependence, however, is removed by fixing the gauge through the requirement that the Wannier states are maximally localized - a natural and logical prerequisite for our real space approach. Since the unitary transformation required to obtain maximally localized Wannier functions is implemented in publicly available software \cite{MOSTOFI20142309}, this structural gauge ambiguity is easily circumvented.

There are, of course, important limitations, even for the proposed method. Our approach cannot be used to discern between these two mechanisms where both are appreciable at a given separation between the lattice sites. This is particularly evident at harmonic orders below the minimum band gap, as depicted in Figs. \ref{fig:2} (c) and (d). It will be interesting to use our approach to analyze the separation-resolved high harmonic emission involving Bragg reflections and from the regions where several conduction bands approach each other, Landau-Dykhne-Zener tunnelling between the bands becomes significant, and the dynamics of the HHG process becomes complex \cite{hhgFromBEiS}. A real-space perspective may offer insight into both the interband tunnelling process and the interplay of different HHG mechanisms in the congested regions of the Brillouin zone \cite{newAlvaro}.

Our approach provides several advantages over previous approaches \cite{Yue:22,PhysRevB.98.235202}: there is no additional computational cost, it is independent of the gauge used to describe the interaction of a system with an external field, and it provides an intuitive and clear picture regarding the mechanisms leading to intra- and interband HHG emission. 

While solid-state HHG has been widely described using models formulated in reciprocal space, the advantages of the Wannier basis have been investigated in numerous works \cite{WQC,intrabandWannier,wannierHHGLewenstein,PhysRevB.100.195201,myFirstHHGSolid}, leading to valuable discoveries and insight regarding displaced tunnelling and electron-hole recombination \cite{WQC,wannierHHGLewenstein}, the spectral characteristics of intraband HHG emission \cite{intrabandWannier}, and the origin of empirically required ultrafast dephasing times in solid-state HHG \cite{firstHHGSolid}. Importantly, the use of maximally localised Wannier functions to describe solid-state HHG has been shown to be an efficient means for circumventing the issues pertaining to the random phase of the dipole couplings in the first Brillouin zone \cite{PhysRevB.100.195201}. In this context, our work reveals yet another advantage of the Wannier basis for describing solid-state HHG, which will be valuable for the advancement of attosecond science in condensed matter.

G. G. Brown acknowledges funding from the European Union’s Horizon 2020 research and innovation programme under grant agreement No. 899794 (Optologic). M. Ivanov  acknowledges funding from the SFB 1477 ``Light Matter Interaction at Interfaces" project number 441234705. \'{A}.J.G. acknowledges funding from the European Union’s Horizon 2020 research and innovation programme under the Marie Skłodowska-Curie grant agreement no. 101028938. R. E. F. Silva acknowledges support from the fellowship LCF/BQ/PR21/11840008 from ``La Caixa” Foundation (ID 100010434

We thank A. Marini, H. Gross, and A. Leitenstorfer for exceptionally useful comments.

The authors declare no conflicts of interest.

Data underlying the results presented in this paper are not publicly available at this time but may be obtained from the authors upon reason- able request.

\bibliography{apssamp}

\end{document}